\documentclass[conference]{IEEEtran}
\usepackage{array}
\usepackage{graphicx}
\usepackage{url}
\usepackage{color}
\usepackage{soul}

\ifCLASSINFOpdf

\else

\fi

\definecolor{myorange}{rgb}{1,0.5,0} 
\sethlcolor{myorange}
\begin{document}

\title{When Machine Learning Meets Vulnerability Discovery: Challenges and Lessons Learned}

\author{\IEEEauthorblockN{Sima Arasteh}
	\IEEEauthorblockA{University of Southern California\\
		arasteh@usc.edu}
	
	\and
	\IEEEauthorblockN{Christophe Hauser\\ Dartmouth College }
	\IEEEauthorblockA{
		christophe.hauser@dartmouth.edu}}

\IEEEoverridecommandlockouts
\makeatletter\def\@IEEEpubidpullup{6.5\baselineskip}\makeatother
\IEEEpubid{\parbox{\columnwidth}{
		Annual Computer Security Applications Conference (ACSAC) 2024\\
		December 9-13, 2024, Hawaii, USA\\
		ISBN 978-1-4503-9999-9\\
		https://dx.doi.org/10.1145/xxxx.xxxx\\
		www.acsac.org
}
\hspace{\columnsep}\makebox[\columnwidth]{}}

\maketitle

\begin{abstract}
In recent years, machine learning has demonstrated impressive results in various fields, including software vulnerability detection. Nonetheless, using machine learning to identify software vulnerabilities presents new challenges, especially regarding the scale of data involved, which was not a factor in traditional methods. Consequently, in spite of the rise of new machine-learning-based approaches in that space, important shortcomings persist regarding their evaluation. First, researchers often fail to provide concrete statistics about their training datasets, such as the number of samples for each type of vulnerability. Moreover, many methods rely on training with semantically similar functions rather than directly on vulnerable programs. This leads to uncertainty about the suitability of the datasets currently used for training. Secondly, the choice of a model and the level of granularity at which models are trained also affect the effectiveness of such vulnerability discovery approaches.

In this paper, we explore the challenges of applying machine learning to vulnerability discovery. We also share insights from our two previous research papers, Bin2vec and BinHunter, which could enhance future research in this field.
\end{abstract}

\IEEEpeerreviewmaketitle

\section{Introduction}
\label{sec:introduction}
Regardless of the method used, a primary challenge in leveraging machine learning to discover software vulnerabilities is obtaining an appropriate training dataset. Many researchers~\cite{feng2016scalable, ding2019asm2vec, xu2017neural, wang2022jtrans} use real-world data, but there are significant issues concerning the data collected. Firstly, detailed statistical information about the data is often missing, such as the number of samples for each vulnerability type or the scope (inter or intra-procedural) of vulnerable samples. Secondly, the diversity of programs used in the training dataset is often limited. As a result of this, models often end up training on semantically similar functions, which introduces bias, rather than on semantically different but similarly vulnerable ones. In Section~\ref{sec:datasets}, we delve into how these issues can impact the quality of the training data.

Researchers have developed various techniques to apply machine learning to discover software vulnerabilities~\cite{ji2021buggraph, ding2019asm2vec, feng2016scalable, chandramohan2018vulseeker, gao2018vulseeker, wang2022jtrans, xu2017neural, tian2020bvdetector, zhou2019devign, lee2017learning, schaad2022deep, chakraborty2021deep, li2021sysevr, eschweiler2016discovre}. However, due to the inherent code structure, utilizing machine learning for this purpose introduces unique challenges. For instance, capturing control and data flow, two crucial properties of program behavior may be essential for some models. Moreover, since vulnerabilities only affect parts of a program, capturing the appropriate level of granularity is important. The choice of an underlying model and its capacity to comprehend program semantics introduces further challenges. In Section~\ref{sec:methods}, we discuss different approaches and techniques to apply machine learning to vulnerability discovery tasks. 

This paper highlights the challenges of leveraging machine learning for vulnerability discovery and points out what aspects of evaluation are missing in the literature. This paper also examines the limitations with current vulnerability discovery techniques from three distinct angles: the datasets used for training and evaluation, the methods of representation learning, and the granularity of learning and detection. In Section~\ref{sec:background}, we summarize our two previous approaches, Bin2vec and BinHunter, both of which focus on binary-level vulnerability discovery. Section~\ref{sec:datasets} reviews all existing training and testing datasets used for vulnerability discovery, along with their limitations. Section~\ref{sec:methods} describes various techniques in terms of representation learning and detection granularity level, along with the limitations of each technique. Additionally, we describe the insights gained from our previous models. Finally,  Section~\ref{sec:baseline}  addresses the challenges associated with selecting and setting up baselines in this domain.

\section{Background}
\label{sec:background}

We have developed two vulnerability discovery approaches, Bin2vec~\cite{arakelyan2021bin2vec} and BinHunter~\cite{arasteh2024:acsac}. In this section, we briefly discuss both. 

\subsection{Bin2vec}
\label{sec:background:bin2vec}
Bin2vec is a binary-level classifier designed to identify vulnerable and patched (non-vulnerable) functions across different types of vulnerabilities. It utilizes a novel graph representation model based on the control flow graph (CFG) of functions, employing graph convolutional neural networks (GCNs) to learn these representations. During the training phase, Bin2vec first generates the CFG for each function, both vulnerable and patched. It then constructs an abstract syntax tree (AST) for each basic block by parsing their intermediate representation (IR). These ASTs are integrated based on the topology of the basic blocks within the CFG. In Bin2vec's graph representation, each node corresponds to an operand or operation in VEX IR~\cite{angrvex}, and the edges represent the structural connections within the AST. For vulnerability discovery, Bin2vec uses the Juliet test suite~\cite{meade2012juliet} as a training and testing dataset. 

\subsection{BinHunter}
\label{sec:background:BinHunter}
BinHunter is a binary classifier designed to train vulnerable and patched binaries for different types of vulnerabilities. It utilizes a specialized graph representation derived from a subgraph of the program dependence graph (PDG) and employs graph convolutional neural networks (GCNs) to detect vulnerabilities. The process begins with BinHunter generating PDGs for both vulnerable and patched programs, from which it then extracts relevant subgraphs. During the training phase, BinHunter identifies function locations associated with vulnerabilities using debug information to form these PDG subgraphs. In contrast, during the testing phase, it employs slicing techniques based on calls to external functions to create subgraphs. Once the subgraph is extracted, BinHunter constructs a graph representation from it, where each node corresponds to operands and operations in the intermediate representation (IR), and the edges denote data dependencies.
For more detailed information on these methods, we refer the reader to~\cite{arakelyan2021bin2vec} and~\cite{arasteh2024:acsac}.

\section{Vulnerability Dataset}
\label{sec:datasets}

Collecting a large number of vulnerable programs is essential for the training phase. Gathering a substantial number of binary programs requires significant effort during the compilation process. Additionally, while a large sample size is necessary for training, it alone does not guarantee the quality of a dataset. Researchers often report only the quantity of data used in their training processes without providing additional statistics~\cite{wang2022jtrans, feng2016scalable}. In this section, we discuss all the essential features of an ideal dataset and identify some threats to existing training datasets.

\subsection{Existing Datasets For Training}
\label{sec:datasets:training}
Compiling a large number of vulnerable binary programs for training is arduous. The Juliet test suite, a synthetic dataset developed by the National Security Agency (NSA) is designed to evaluate the efficacy of static analysis tools in detecting vulnerabilities within code. It encompasses a wide array of C/C++ and Java programs that contain both vulnerable and non-vulnerable versions of each test case. In each sample, macros define the vulnerable and non-vulnerable code. This dataset covers over a hundred types of vulnerabilities and includes a large number per type of vulnerability. There are some advantages and limitations associated with this dataset. The Juliet test suite has several advantageous features:

\begin{itemize}
    \item First, it is an easy-to-compile dataset and it covers a large number of vulnerable and non-vulnerable programs. 
    \item The programs in the Juliet test suite exhibit a diverse range of semantics.
    \item It covers both intra-procedural and inter-procedural vulnerabilities.
    \item It covers a diverse range of data and control flow complexity in the programs. The Juliet test suite assigns a number to each program that defines a level of data and control flow complexity~\cite{meade2012juliet}.
\end{itemize}

\begin{table*}[ht]
\centering
\caption{Statistical Overview of the Juliet Test Suite Dataset: This table details the count of samples within various complexity and procedural contexts for each CWE category. Columns include intra-procedural samples showing vulnerabilities within single functions, inter-procedural samples across multiple functions, and samples with data flow (DF) and control flow (CF) complexities. The final column combines counts of samples with both data and control flow complexities. }
\label{tab:juliet_statistic}
\renewcommand{\arraystretch}{1.2} 
\begin{tabular}{|>{\centering\arraybackslash}m{0.1\linewidth}|
                >{\centering\arraybackslash}m{0.1\linewidth}|
                >{\centering\arraybackslash}m{0.1\linewidth}|
                >{\centering\arraybackslash}m{0.1\linewidth}|
                >{\centering\arraybackslash}m{0.1\linewidth}|
                >{\centering\arraybackslash}m{0.1\linewidth}|} 
\hline
\textbf{CWE-ID} & \textbf{\# intra-procedural} & \textbf{\# inter-procedural} & \textbf{\# DF complexity} & \textbf{\# CF complexity} & \textbf{\# DF and CF complexity} \\ 
\hline
CWE-121 & 2619 & 5252 & 4644 & 1179 & 342 \\ \hline
CWE-122 & 2878 & 6752 & 5604 & 1500 & 366 \\ \hline
CWE-124 & 1038 & 2274 & 1956 & 492 & 138 \\ \hline
CWE-126 & 786 & 1506 & 1284 & 378  & 90  \\ \hline
CWE-127 & 1038 & 2274 & 1956 & 492 & 138 \\ \hline
CWE-190 & 2034 & 4410 & 3636 & 1080 & 270 \\ \hline
CWE-191 & 1551 & 3171 & 2670  & 828 & 207 \\ \hline
CWE-369 & 414 & 1062 & 828 & 216 & 54 \\ \hline
CWE-400 & 345 & 885 & 690 & 180 & 45 \\ \hline
CWE-401 & 890 & 1814 & 1452 & 462 & 102 \\ \hline
CWE-415 & 460 & 1184  & 920 & 240 & 60  \\ \hline
CWE-416 & 378 & 142 & 140  & 189 & 0 \\ \hline
CWE-590 & 1474 & 2747 & 2546 & 603 & 201 \\ \hline
CWE-680 & 276 & 660 & 552 & 144 & 36 \\ \hline
CWE-690 & 460 & 1100 & 920 & 240 & 60 \\ \hline
CWE-762 & 1702 & 4390 & 3404 & 888 & 222 \\ \hline
\end{tabular}
\end{table*}

\begin{table*}
\centering
\caption{Summary of existing methods, their training and testing dataset for vulnerability discovery.}
\label{tab:approches}
\renewcommand{\arraystretch}{1.5}
\scalebox{0.85}{
\begin{tabular}{|>{\centering\hspace{0pt}}m{0.15\linewidth}|>{\centering\hspace{0pt}}m{0.36\linewidth}|>{\centering\hspace{0pt}}m{0.3\linewidth}|>{\centering\hspace{0pt}}m{0.135\linewidth}|>{\centering\arraybackslash\hspace{0pt}}m{0.1\linewidth}|} 
\hline
Approach~  & Training dataset                                                                                                                                                                                                                                                             & Testing dataset                                                                                                                                                                                      & Application                                               & Availability~  \\ 
\hline
Genius~\cite{feng2016scalable}    & Contains functions in BusyBox (v1.21 and v1.20), OpenSSL (v1.0.1f and v1.0.1a) and Coreutils \\(v6.5 and v6.7) Compiled for three different architectures (x86, ARM, MIPS) using three compiler versions (gcc v4.6.2/v4.8.1 and clang v3.0) and four optimization levels & Contains 154 \par{}vulnerable functions~                                                                                                                                                             & Binary Similarity Detection, Vulnerability Discovery & -              \\ 
\hline
discovRE~\cite{eschweiler2016discovre}   & Contains functions compiled from BitDHT, GnuPG, ImageMagick, LAME, OpenCV, SQLite, and stunnel compiled for various compilers, optimization options, and for different CPU architectures and operating systems                                                   & Limited functions in OpenSSL binaries~ (\textit{dtls1\_process\_heartbeat,} \textit{tls1\_process\_heartbeat, ssl\_cipher\_list\_to\_bytes}) to find Heartbleed and POODLE vulnerabilities & Binary Similarity Detection, Vulnerability Discovery & -              \\ 
\hline
VDiscover~\cite{grieco2016toward} & Debian programs                                                                                                                                                                                                                                                              & Debian programs                                                                                                                                                                                      & Vulnerability Discovery                                   & -              \\ 
\hline
Gemini~\cite{xu2017neural}     & Functions in openssl (version 1.0.1f and 1.0.1u) compiled for three different architectures and 4 optimization levels~                                                                                                                                                 & Contains 154 vulnerable functions                                                                                                                                                                    & Binary Similarity Detection, Vulnerability Discovery & -              \\ 
\hline
Asm2vec~\cite{ding2019asm2vec}     & Semantically similar and benign functions in BusyBox, CoreUtils, Libgmp, ImageMagick, Libcurl, LibTomCrypt, OpenSSL, SQLite, zlib and PuTTYgen ~                                                                                                                                                 & 8 number of CVEs from~\cite{david2016statistical}                                                                                                                                                                   & Binary Similarity Detection, Vulnerability Discovery & -              \\ 
\hline
BVdetector~\cite{tian2020bvdetector} & SARD                                                                                                                                                                                                                                                                         & SARD                                                                                                                                                                                                 & Vulnerability Discovery                                   & Public         \\ 
\hline
Trex~\cite{pei2020trex}       & A real world dataset of benign programs contains 1,472,066 functions from 13 projects                                                                                                                                                                                  & 16 CVEs extracted from openssl and busybox                                                                                                                                                           & Binary Similarity Detection,\par{}Vulnerability Discovery & Public         \\ 
\hline
BugGraph~\cite{ji2021buggraph}   & 493,841 number of similar and benign functions compiled from 2 software (SNNS-4.2, PostgreSQL-7.2) and six different compilers four optimization levels for x86                                                                                                   & 42 CVEs contains 218 vulnerable functions                                                                                                                                                       & Binary Similarity Detection, Vulnerability Discovery & -              \\ 
\hline
Bin2vec~\cite{arakelyan2021bin2vec}    & Juliet Test Suite                                                                                                                                                                                                                                                            & Juliet Test Suite                                                                                                                                                                                    & Binary Similarity Detection,\par{}Vulnerability Discovery & Public         \\ 
\hline
Jtrans~\cite{wang2022jtrans}   & real-world and benign programs~extracted  from ArchLinux repositories from 1,612 projects. The name of the dataset id BinaryCorp                                                                                                                                   & A collection of~

~8 CVES                                                                                                                                                                            & Binary Similarity Detection, Vulnerability Discovery & Public         \\ 
\hline
VulHawk~\cite{luo2023vulhawk}    & A dataset of real-world and begin programs from 10 programs compiled for different optimization levels                                                                                                                                                                 & 12 CVEs from openssl and curl contains 93 vulnerable functions                                                                                                                                 & Binary Similarity Detection, Vulnerability Discovery & Public         \\ 
\hline
VulCatch~\cite{chukkol2024vulcatch}   & SARD                                                                                                                                                                                                                                                                         & 7 CVEs                                                                                                                                                                                               & Vulnerability Discovery                                   & -              \\ 
\hline
CEBin~\cite{wang2024cebin}     & use existing datasets, BinaryCorp of Jtrans , CisCo~\cite{marcelli2022machine} and Trex                                                                                                                                                                                                                 & 187 CVEs from 5 projects. curl, vim, libpng, openssh and openssh-portable                                                                                                                      & Binary Similarity Detection, Vulnerability Discovery & Public         \\ 
\hline
BinHunter~\cite{arasteh2024:acsac}  & Juliet Test Suite                                                                                                                                                                                                                                                         & Juliet test suite, 24 CVEs extracted from Debian packages                                                                                                                                       & Vulnerability Discovery                                   & Public         \\
\hline
\end{tabular}
}
\end{table*}

\begin{table}
\centering
\caption{Existing Binary datasets used for different applications}
\label{tab:binary datasets}
\renewcommand{\arraystretch}{1.5}
\scalebox{0.9}{
\begin{tabular}{|>{\centering\hspace{0pt}}m{0.273\linewidth}|>{\centering\hspace{0pt}}m{0.452\linewidth}|>{\centering\arraybackslash\hspace{0pt}}m{0.192\linewidth}|} 
\hline
Dataset    & Application                                           & Availability  \\ 
\hline
ASSEMBLAGE~\cite{liu2024assemblage} & Binary similarity detection,\par{}compiler provenance & Public        \\ 
\hline
BinBench~\cite{console2023binbench}   & Binary similarity detection,\par{}Compiler Provenance & Public        \\ 
\hline
BinaryCorp~\cite{wang2022jtrans} & Binary similarity detection                           & Public        \\ 
\hline
PunStrip~\cite{patrick2020probabilistic}   & Function name recovery                                & -             \\ 
\hline
Nero~\cite{david2020neural}      & Function name recovery                                & Public        \\ 
\hline
BinKit~\cite{kim2022revisiting}     & Binary similarity detection                           & Public        \\ 
\hline

\end{tabular}
}
\end{table}

Table~\ref{tab:juliet_statistic} presents statistics for the Juliet test suite dataset, detailing the number of intra-procedural and inter-procedural programs by type of vulnerability, along with the number of programs addressing data and control flow complexity. Intra-procedural programs consist of those where vulnerabilities are confined to a single function, whereas inter-procedural programs include those where vulnerabilities span multiple functions. 

Although Juliet has the qualities of an effective training dataset, it has some limitations. The vulnerabilities are manually injected and the programs are constructed artificially. Moreover, the functions in the Juliet test suite are small. Despite these limitations, we employed the Juliet test suite in both the training and testing phases of our models. Our goal was to assess our methods' effectiveness in detecting vulnerabilities within real-world software. To achieve this, we trained our models using the Juliet dataset and tested them on actual vulnerable programs. Conducting this experiment presented several challenges.

\begin{itemize}
 \item \textbf{Challenge-1:} The first challenge was finding a suitable dataset of real-world vulnerable programs that met our requirements. Unfortunately, no existing dataset of real-world vulnerable programs met our expectations. 
    \begin{itemize} 
        \item \textbf{Mitigation:} To overcome this, we created a new dataset using historical vulnerability data from Debian packages. Assembling a dataset from real-world data requires significant effort, even though Debian provides tools for automatic package construction. We built the packages from the source code for both vulnerable and patched versions. Building this dataset presented several challenges. \textit{First}, many vulnerable packages are old, and providing an appropriate environment to build such packages can be cumbersome. \textit{Second}, to build the packages, we start with the package source code that contains a fix to a specific CVE, then we leverage quilt\footnote{https://wiki.debian.org/UsingQuilt} to apply or remove the patch. In some cases, quilt produces some errors while reverting the patch. \textit{The next challenge} arises in a few instances where, after applying or reversing the patch with quilt, the diff lines containing the patch do not align with the corresponding lines in the original source code.
    \end{itemize}
    \item \textbf{Challenge-2:} The second challenge involves addressing the following question: \textit{Is the knowledge gained from Juliet transferable to real-world programs?} This question forms the basis of a crucial experiment we need to conduct. 
        \begin{itemize}
            \item \textbf{Mitigation:} To assess the feasibility of this experiment, we trained our classifier only on vulnerable and patched program locations. In the training phase, we can assume access to the source code and debug symbols, whereas, in the testing phase, we cannot make this assumption. However, to evaluate the feasibility of transferring knowledge from Juliet to real-world programs, we first tested our classifier on the exact bug locations in the real-world programs of the test set. If the classifier can correctly identify both the bug and patched locations, then it provides a basis for proceeding with further experiments. Note that the slicing technique we used in BinHunter paved the way to make this experiment feasible. 
        \end{itemize}

\end{itemize}

\textbf{Threats to Validity:} In our experiments, as documented in the paper~\cite{arasteh2024:acsac}, we tested only 24 vulnerable functions, and BinHunter successfully detected 17 of them. However, training on small-sized functions such as Juliet is not scalable on a large scale. Consequently, there is a need for a more comprehensive dataset for training.

Due to the artificial nature of the Juliet dataset, researchers~\cite{feng2016scalable, xu2017neural, ding2019asm2vec, wang2022jtrans} usually gather and build a dataset from real-world programs. However, there are some challenges and concerns regarding these datasets.

\begin{itemize}
    \item Compiling a large number of vulnerable and patched real-world programs is arduous.
    \item Most real-world datasets are not publicly available. As Table~\ref{tab:approches} suggests among real-world datasets only 5 of them are publicly available. Unfortunately, some datasets, such as Genius~\cite{feng2016scalable}, are not available despite being claimed as such in the paper.
    \item There is a lack of statistical information (e.g., number of samples per CWE) about the existing datasets.
    \item The majority of the training datasets~\cite{feng2016scalable, wang2022jtrans} consist of benign functions with similar semantics, such as functions compiled for various optimization levels and architectures. The classifier is trained on semantically similar functions and tested on a limited number of vulnerable programs. Table~\ref{tab:approches} lists the available training and testing datasets used for vulnerability discovery from the current literature. As suggested by this table, many approaches train on semantically similar and benign functions and test on a limited number of vulnerable programs. For instance, Genius, discovRE, Gemini, and Asm2vec contain functions of three, seven, two, and ten projects compiled for different architectures, optimization levels, and compilers respectively.  
     \item The datasets used for training suffer from a lack of program diversity. As Table~\ref{tab:approches} shows Jtrans contains the most diverse programs. It contains 1,612 number of projects, while others such as Genius, BugGraph, Asm2vec only cover a few number of projects. If the classifier is trained on semantically identical programs, it is unclear whether it is learning to identify vulnerabilities or merely capturing program semantics.  
\end{itemize}

We also gather existing binary program datasets used for different purposes in Table~\ref{tab:binary datasets}. Many binary collection datasets, such as Assemblage~\cite{liu2024assemblage}, BinBench~\cite{console2023binbench}, Binarycorp~\cite{wang2022jtrans}, and BinKit~\cite{kim2022revisiting}, are specifically designed for binary similarity detection. However, some methods like Jtrans [29] utilize these datasets to train models for vulnerability discovery tasks. In such instances, it remains uncertain to what extent a classifier trained on semantic binary functions can identify various types of vulnerabilities. As Table~\ref{tab:approches} and~\ref{tab:binary datasets} suggests, existing binary program collections are not suitable for vulnerability discovery tasks, and we need a specific and comprehensive dataset for vulnerability discovery.
 
\textbf{Threats To Validity:} 
Lack of semantic diversity in the programs of the training and testing datasets can potentially lead to classifier overfitting. Furthermore, most existing models are not trained on both vulnerable and patched programs; rather, they are trained on semantically similar programs. Consequently, the effectiveness of these models in detecting vulnerabilities on a large scale remains uncertain.

\subsection{Features of an Ideal Dataset}
\label{sec:datasets:ideal}
Due to the aforementioned problems with existing vulnerability discovery datasets, there is a lack of an ideal dataset with these features for training.

\begin{enumerate}
    \item A large number of samples for each type of vulnerability.
    \item Semantically diverse programs in both training and testing phases. 
    \item Public availability.
    \item Coverage of both inter-procedural and intra-procedural vulnerabilities. 
\end{enumerate}

\subsection{Suggestions}
\label{sec:datasets:suggestions}
There is a significant gap in evaluating the quality of existing training datasets. A possible approach is to train models on their respective datasets and test them on a completely different set, as we proposed in~\cite{arasteh2024:acsac}. The majority of existing models were trained on programs that are semantically similar~\cite{feng2016scalable, wang2022jtrans} rather than directly on vulnerable ones. These studies aimed to detect similar functions and included one experiment to assess the model's capability to discover vulnerabilities.

There is a need to create a dataset for training that can meet the features of an ideal dataset for vulnerability discovery. Also, there should be a study to train different models using this ideal dataset and test on a completely different vulnerability dataset. The functions in the training and testing set should be selected from completely different projects.

\section{Methodologies}
\label{sec:methods}

The variations in existing deep-learning-based methodologies arise from the responses to the following questions:

\begin{itemize}
    \item \textit{What granularity level should be chosen for training and testing?}
    \item \textit{Which representation learning approach should be selected?}
    \item \textit{Which code properties should be captured and what methods should be used to capture them?}
\end{itemize}

\subsection{Granularity Level}
\label{sec:methods:granularity}
The majority of existing models treat function similarity detection and vulnerability discovery as the same problem~\cite{ding2019asm2vec, wang2022jtrans}. They train on semantically similar functions and convert the entire function into a vector representation for the neural network. During the testing phase, they also detect vulnerabilities at the function level. On the other hand, some techniques train and test at finer granularity levels. Bin2vec uses function-level granularity, while BinHunter leverages a slicing technique to train and test only on code locations within a function that is relevant to vulnerability. Both Bin2vec and BinHunter were trained on the Juliet test suite dataset. We evaluated both Bin2vec and BinHunter on the Juliet test suite dataset and vulnerable real-world programs derived from Debian packages.  Our experimental results proved that among 24 vulnerable real-world programs, Bin2vec could only detect 2 bugs while BinHunter could detect 17 of them. The slicing technique could help transfer knowledge from small functions of Juliet to big functions in real-world programs. We explain this behavior by the fact that BinHunter only relies on function subgraphs related to vulnerable code.

\textbf{Threats to Validity:}
Relying on function-level granularity for training raises some concerns. When learning at the function granularity level, the classifier primarily learns the function's semantics instead of vulnerability patterns. This issue is particularly unclear for large functions, where it is uncertain whether the classifier has learned vulnerability patterns or merely the function semantics.
While relying on a finer granularity level, as we did with BinHunter, can address this issue, it may also introduce new challenges. Extracting the location of vulnerabilities within functions during training is straightforward, but it becomes challenging during testing due to the lack of access to source code and debug symbols. The reliability of these models highly depends on the robustness of the slicing techniques used during testing.
\subsection{Representation Learning} \label{sec:methods:Representation} Representation learning transforms software code into vector embeddings, enabling neural networks to capture the underlying patterns and features characteristic of software. This approach aids in identifying vulnerabilities by representing code in a more structured form that highlights patterns indicative of security issues. Researchers have applied various techniques to leverage this type of learning for vulnerability discovery, enabling models to efficiently recognize and analyze vulnerable code.
Some researchers treat binary functions as text and apply natural language processing (NLP) techniques to model vulnerable programs. This approach represents code as a sequence of tokens, similar to how words are treated in NLP tasks, utilizing methods like word embeddings and recurrent neural networks (RNNs). In contrast, models like Bin2vec and BinHunter use graph-based representations, where code is modeled as a graph of nodes (e.g., variables) and edges (representing relationships between them). This graph structure is better suited for capturing complex interactions and relationships within the code.

While NLP-based models have their strengths, they face limitations in capturing the intricate relationships between code variables and structures. These models rely on sequence-based attention mechanisms, which might overlook the hierarchical and relational nature of code. This makes graph-based models, such as Bin2vec and BinHunter, more effective for capturing the complex relationships within the software, leading to better vulnerability detection.
\subsection{Code Properties to Capture} \label{sec:methods:Properties} Two fundamental properties of programs are control and data flow. Control flow refers to the sequence in which instructions or statements are executed in a program, describing the paths that the program takes based on conditions like loops, branches, or function calls. Data flow, on the other hand, tracks how data is passed between variables, functions, or memory locations during program execution. These two code properties are crucial for vulnerability discovery because control flow helps identify logical errors or flaws in the execution path, while data flow reveals how malicious inputs or unintended data manipulations can propagate, leading to security issues such as buffer overflows, information leaks, or injection attacks. Analyzing program semantics through the lens of both control flow and data flow enables a deeper understanding of how vulnerabilities may arise from improper handling of logic or data in a program.

Our Bin2vec model captures control flow, while BinHunter captures both control and data dependencies. As mentioned in Section~\ref{sec:datasets}, programs in the Juliet test suite dataset are labeled based on the complexity of their data and control flows. Analyzing the misclassified samples from our first model, Bin2vec, revealed that the majority of these misclassifications correspond to programs with complex data flows. This analysis revealed the need to capture data flows in our new model, BinHunter. However, generating data flows at the binary level, especially capturing inter-procedural data flow between functions, is very difficult and comes with its own limitations. Our ablation study, which assessed the impact of data and control flow factors on performance, showed that incorporating both code properties can enhance detection effectiveness. See the results in Tables 3 and 4 in BinHunter~\cite{arasteh2024:acsac}.

Capturing either control or data flow had some implementation challenges in practice. To capture control flow in Bin2vec and the Program dependence graph (PDG)  in BinHunter, we used angr~\cite{shoshitaishvili2016state} and Ghidra~\cite{ghidra} respectively. Both are powerful reverse engineering tools commonly used for analyzing binary programs. Ghidra, developed by the NSA, is a comprehensive software reverse engineering framework that offers disassembly, decompilation, and debugging capabilities. Angr, on the other hand, is a Python-based platform focused on symbolic execution, vulnerability analysis, and program exploration. There were both advantages and disadvantages to using these tools when implementing our previous models. Angr supports both intra-procedural and inter-procedural data flows, while Ghidra is limited to supporting only intra-procedural data flows. Moreover, Ghidra lacks extensive code examples for its API usage, which complicates its application. For example, implementing a PDG required us to consult with the developers for a deeper understanding of the source code.
While developing BinHunter, we used both angr and Ghidra to generate data dependencies. After analyzing the accuracy of generated data flows from both angr and Ghidra by mapping source code lines to binary memory offsets using debug symbols, we decided to use Ghidra in our BinHunter tool.

\section{Challenges of Baseline Comparison:}
\label{sec:baseline}
To select our baselines, we consider several factors: the inclusion of experiments related to vulnerability discovery, the use of open-source tools, and relevance to our specific problem. To find appropriate baselines and conduct a fair comparison, we faced these challenges.

\begin{itemize}
    \item \textbf{Challenge-1:} All of our baselines were bug search engines. They compare a target program (e.g. a vulnerable function) against functions in their repository. They rank repository functions based on the similarity to the target function and use some evaluation metrics on top n retrieved functions. However, our previous models are binary classifiers. To evaluate our model, we compare their output to the ground truth labels. Finding fair evaluation metrics and comparisons was a big challenge. 
    \begin{itemize}
        \item \textbf{Mitigation:} To tackle this challenge, we employ the logic used in binary classification. In binary classification, if the probability of a sample exceeds 0.5, it is classified as class A; otherwise, it is classified as class B. We apply this principle to bug search engine models. For example, consider a scenario where the original model evaluates the top N retrieved functions. If at least half of these top N functions share the same label as the target function—such as being vulnerable if the target function is also vulnerable—then the detection is considered correct.
    \end{itemize}
    \item \textbf{Challenge-2:} The second challenge involved some baselines using tools like IDA Pro, which require a license and are costly.
    \begin{itemize}
        \item \textbf{Mitigation:} We reached out to Hexray to request the educational version of the IDA-Pro license. The license is only valid for a short period.
    \end{itemize}
    \item \textbf{Challenge-3:} Adapting certain baselines to our dataset was challenging due to variations in implementation details. For example, two of our baselines, Jtrans and Gneuis, used in BinHunter, required modifications to adapt to our dataset.
    \begin{itemize}
        \item \textbf{Mitigation:} To address this challenge, we needed to understand the implementation details of the baselines and modify them while preserving their overall structure.
    \end{itemize}
\end{itemize}

\section{Conclusion}

We have developed two binary vulnerability discovery methods, Bin2vec and BinHunter. Our experience with these models has shown that careful consideration is crucial when selecting benchmarks for training. Most existing methods train on semantically similar programs, and there is a lack of comprehensive datasets that contain diverse types of vulnerabilities for training. An ideal dataset should include a diverse range of programs, encompass various types of vulnerabilities, and cover both vulnerable and patched programs. Additionally, when selecting representation learning methods and determining the granularity level for learning and detection, researchers should consider that each method has its own advantages and disadvantages, and these should be carefully evaluated during the selection process. Finally, in selecting baselines, it is important to note that many are not freely available or require proprietary licenses, and sometimes modifications are needed to accommodate minor changes in the problem setting. Careful selection and adaptation of these elements are essential for applying machine learning on vulnerability discovery.


\bibliography{src/references}
\bibliographystyle{plain}

\end{document}